\documentstyle[mncite]{mn}
\title[Distance to $\alpha$ Persei and the Pleiades]
      {New distance measurements to the $\alpha$~Persei and Pleiades clusters}

\author[M.A. O'Dell, M.A Hendry and A. Collier Cameron]
        {M.A.O'Dell, M.A. Hendry and A. Collier Cameron\\
	 Astronomy Centre, University of Sussex,
         Falmer, Brighton BN1 9QH, UK.}

\date{Accepted ------. Received ------; in original form \today}

\begin{document}

\maketitle

\input latex_macros.tex

\begin{abstract}

We apply the new distance measuring technique of
\scite{hendry93} to the Pleiades and $\alpha$ Persei clusters.
The method relies on knowledge of the periods, rotational velocities
and angular diameters of a small sample of the late-type stars
within each cluster.
The angular diameters of these stars were found by recalibrating
the Barnes-Evans surface brightness
relation for an improved calibration sample in the spectral
range $-0.15 \leq (B - V)_0\leq1.35$.
Applying our new distance method we derive a distance
modulus, $(m - M)_v$ = 5.60 mag. for the Pleiades and
$(m - M)_v$ = 6.35 mag. for $\alpha$ Persei.
Our determination of the Pleiades distance is in close
agreement with previous zero-age main-sequence fitting techniques,
whereas our distance for $\alpha$ Persei appears substantially greater.
In response to the greater distance for $\alpha$ Persei we apply
the relatively recent semi-empirical main-sequence fitting
method of \scite{vandp89} and confirm our findings,
obtaining a distance of $(m - M)_v$ = 6.25 mag.
\end{abstract}


\noindent

\section{INTRODUCTION}
\label{sec:intro}

The accurate determination of the distances to nearby open clusters is
a longstanding problem, the solution of which would greatly improve our
understanding of the Universe. Of direct benefit would be the precise
determination of intrinsic stellar luminosities, the study of local
galactic structure and not least the calibration of the extragalactic
distance scale.

In our preceding paper, \scite{hendry93} (hereafter HOC) we
described a new technique for estimating the distance to nearby young
open clusters via a knowledge of the axial rotation periods of their
late-type members. We showed that for a simulated cluster sample it was
possible to model the axial rotation period,
projected equatorial velocity,
$v{\sin}i$, and angular diameter for a predetermined number of cluster
members, so as to generate the distribution of their projected cluster
distances, $D{\sin}i$.
By ordering these projected distances we showed how one may construct an
estimator for the true cluster distance whose accuracy lies between
${\sim}$3 and 7 per cent at the 1 ${\sigma}$ level, depending on the
number of sampled stars and the details of our model assumptions.
An important feature of the method is that it is subject to a completely
different set of systematic errors and model assumptions than the more
traditional distance estimators such as the ZAMS fitting and moving
cluster methods and therefore acts as an independent check of the local
distance scale.

In this paper we apply this technique to real data samples taken from
the Pleiades and $\alpha$ Persei clusters to infer their true distances and
then compare the results with a recent zero-age main sequence fitting
technique (hereafter ZAMS fitting technique).

It would be worthwhile at this stage to briefly mention previous attempts
to measure the distances to these clusters not least as a background
into the problem of distance measurement but also to place into context
the results of our method.

There exists a variety of techniques that measure the distance to galactic
clusters, although unfortunately many apply only to the very nearest
clusters.
Table \ref{tab:distdata} shows some of the previous distance determinations
for the Pleiades and $\alpha$ Persei.

\begin{table*}
  \centering
  \caption{ Previous determinations of distance for the Pleiades
and $\alpha$ Persei open clusters.}
  \label{tab:distdata}
  \input dist-tab.tex
\end{table*}

We discuss in detail here only the ZAMS fitting technique, which has been
used almost exclusively, being undoubtedly the most preferred as it is based
on non-evolved stars which are believed to be fairly well understood.
The technique was developed primarily by \scite{johnsonh56},
\scite{sandage57}, \scite{johnson57} and \scite{johnsoni58}.
Unfortunately in its standard form it requires a calibrating cluster
(normally the Hyades) whose distance has been derived using other
methods, (e.g from proper motion studies and trigonometric parallax
measurements) and therefore is subject to the errors associated with these
techniques. It also assumes that the ZAMS for all clusters is identical and
that stars resident on the ZAMS are unaffected by differences in their
initial chemical composition.

In an attempt to improve the ZAMS fitting technique by considering
the initial composition of the cluster members
\scite{vandb84} and \scite{vandp89} introduced a
semi-empirical zero-age main-sequence fitting technique employing
computations of model stellar atmospheres.
Their method consists of
fitting a ZAMS locus (defined by a quartic), normalised to the Sun,
to the Colour-Magnitude (C-M) diagram of the cluster whose
distance is to be determined.
The position of this locus on the C-M diagram is
set by the helium and metal abundances of the cluster
members. In addition the locus is constrained by the morphological
profile of the Pleiades ZAMS and chemical composition of the population II
subdwarf Groombridge 1830.
This method appears to be the most rigorous ZAMS fitting technique
available to date: we therefore employ it to check the results of our own
distance method.
The measurement of the Pleiades distance is given in Table \ref{tab:distdata},
\cite{vandb84}. The measurement of the distance to $\alpha$ Persei we
estimate using the method of \scite{vandp89} (see section 4).

A necessary parameter required by our new distance measuring technique is
the angular diameter of each star in our cluster sample.
In HOC we obtained this parameter via the surface brightness relation
of Barnes and Evans \cite{barnes78}. In section \ref{sec:recalib}
we recalibrate the Barnes-Evans relation using the recent results of the
IRAS 12${\mu}m$ measurements quoted in \scite{perrink87}, in conjunction
with 27 calibration stars of Barnes and Evans.
In section \ref{sec:applic} we use our new technique described in HOC to infer
the distances to both clusters.
Finally in sections \ref{sec:zams} and \ref{sec:disc}
we compare our results with those obtained
using the ZAMS fitting technique and discuss their implications.

\section{RECALIBRATION OF THE BARNES-EVANS RELATION}
\label{sec:recalib}

We showed in HOC (equation 4) that the projected cluster distance $D{\sin}i$
for each predetermined cluster member can be given by:
\begin{equation}
\label{eq:dsinival}
   D{\sin}i = 7.660{\times}10^{-3}\; \frac{Pv{\sin}i}{\phi}\; ,
\end{equation}
where for each star, $P$ is the axial rotation period (hours), $v{\sin}i$
is the projected equatorial velocity (\kmsec) and ${\phi}$ is the
angular diameter (milliarcseconds). These units give $D{\sin}i$ in parsecs.

The parameters $P$ and $v{\sin}i$ can be measured directly by observation;
using photometry to measure the period via the rotational modulation
of the stars flux output due to starspots, and spectroscopy to measure the
doppler broadening of its spectral lines due to rotation.

The angular diameter $\phi$ however requires a more indirect approach.
In HOC (section 2) we showed that $\phi$ could be determined using the
semi-empirical relation of \scite{barevans76}. This is a hybrid
combination of photometry with angular diameter measurements found from
lunar occultation and interferometry observations of nearby stars.

As mentioned in HOC, the Barnes-Evans relation is dependent on a
correlation existing between the visual surface brightness of a star and
its angular diameter. In \scite{barevans76} the relation is given as:
\begin{equation}
\label{eq:logTeBEM}
   \log{T_e} + 0.1C = 4.2207 - 0.1{V_0} - 0.5{\log{\phi}}
\end{equation}
Where $T_e$ is the effective temperature, $C$, the bolometric correction
and $V_0$ the unreddened apparent magnitude in the Johnson UBVRI
photometric system. The right-hand side of this equation they define as
the surface brightness parameter, $F_v$, which they show correlates
well with the unreddened colour indices $B - V$, $V - R$ and $R - I$.
The right-hand side of equation \ref{eq:logTeBEM} includes the calibrating
constant, 4.2207, which relies on the solar values of $T_e$, $C$, $v_0$ and
$\phi$. We have determined this constant and find it to be 4.2137: the
difference from the Barnes-Evans value appears due to the choice of the
solar effective temperature which is dependent on current model atmosphere
theory. Our value is consistent with that quoted in \scite{gray92}
(equation 15.2). Equation \ref{eq:logTeBEM} therefore becomes:
\begin{equation}
\label{eq:logTeOHC}
   F_v = \log{T_e} + 0.1C = 4.2137 - 0.1{V_0} - 0.5{\log{\phi}}
\end{equation}

In later papers \scite{barnespar76} and \scite{barnes78} improved the
relationship between $F_v$ and the respective colour indices by adding
further stars to their calibration sample.

They showed that the plots of $F_v$ against $V - R$, $R - I$ and $B - V$
all exhibit linear behaviour over the colour range of the $\alpha$ Persei
and Pleiades G-K dwarfs whose axial rotation periods are known. Moreover they
found that the $V - R$ plot displayed the least scatter of all the
relationships. However as stated in HOC we preferred to use the $F_v$ vs
$B - V$ relation since the effects of interstellar reddening were minimal for
this colour. This is because the slope of this relation more closely
approximates that of the reddening line. \scite{barnes78} give the
following linear equation for the spectral range of our concern:
\begin{equation}
\label{eq:FvBEM}
   F_v = 3.964 - 0.333(B - V)_0, \,\,\, -0.10 \leq (B - V)_0\leq1.35
\end{equation}
where $(B - V)_0$ is the unreddened colour index.

Unfortunately close examination of their $F_v$ vs $B - V$ plot reveals that
only two calibration  stars (excluding the Sun) fall into the spectral
range of our G-K dwarfs. In addition nearly all of the Barnes-Evans
calibration stars that give the above linear relation are giants and not
young main-sequence stars as in our cluster sample.
\scite{popper80} recognised these problems which apply when using the
Barnes-Evans relation to main-sequence stars. This prompted him
to modify the relation by considering a higher order fit to the
Barnes-Evans data.
Similarly we decided that in order to obtain the most accurate estimates
of angular diameter for our stars a modification or recalibration was
necessary.

This was carried out by employing the results of \scite{perrink87} who
determined the angular diameters and radii for 35 G to K type dwarfs using
IRAS 12${\mu}m$ measurements and surface flux estimates computed from
model atmospheres.
Their method (used originally by \scite{gray67}) uses the following
expression to determine angular diameter:
\begin{equation}
\label{eq:phiIRAS}
   \phi = 2 (f_v/{\cal F}_v)^{1/2}
\end{equation}
where $\phi$ is in radians and
$f_v$ is the flux received at the Earth found after converting the
broadband IRAS measurements to a monochromatic flux and ${\cal F}_v$ is the
monochromatic flux leaving the surface of the star.

The program stars of \scite{perrink87} substantially increased the number of
calibration stars within the spectral range of our cluster members, and
allowed us to recalibrate the Barnes-Evans relation.

We decided to take as our calibrating sample the stars of Perrin and Karoji
(hereafter PK stars) and the Barnes-Evans stars (hereafter BEM stars)
whose $(B - V)_0$ lay within
the range $-0.15\, \leq \,(B - V)_0\, \leq \,1.35$.
We then derived $F_v$ for the entire sample using the right-hand side of
equation \ref{eq:logTeOHC}. The angular diameters of the BEM stars were found
directly using lunar occultation and interferometry observations and
are recorded in \scite{barnes78}. The angular diameters of the PK stars
were found from the IRAS 12${\mu}m$ fluxes (equation \ref{eq:phiIRAS}) and are
recorded in \scite{perrink87}. Figure \ref{fig:BEreln} shows the
resulting $F_v$ values plotted against the unreddened colour index,
$(B - V)_0$
\footnote{The PK stars all lie in the solar neighbourhood.
We therefore assumed that the effects of interstellar absorption were
negligible , i.e. $B - V\, \equiv\, (B - V)_0$. The BEM stars colours
are all corrected for reddening.}.
\begin{figure*}
  \centering
  \vbox to100mm{}
  \caption{A recalibration of the Barnes-Evans relation, $F_v$ vs
   $(B - V)_0$.}
  \label{fig:BEreln}
\end{figure*}
A correlation between $F_v$ and $(B - V)_0$ clearly exists and is well
described by a linear fit. Performing an orthogonal regression (i.e.
accounting for errors on both variables) on the calibrating sample we
obtained the following equation:
\begin{equation}
\label{eq:FvOHC}
   F_v = 3.961 - 0.341(B - V)_0, \,\,\, -0.15 \leq (B - V)_0\leq1.35
\end{equation}
This relation replaces equation \ref{eq:FvBEM} for the remainder of this paper.

In addition we examined the PK stars to check that $F_v$ derived from the
right-hand side of equation \ref{eq:logTeOHC} correlated well with that
obtained from the left-hand side. This was to confirm that the angular
diameters obtained using equation \ref{eq:phiIRAS} were consistent with those
of Barnes and Evans (see Figure \ref{fig:fvcorrel}).
\begin{figure}
  \centering
  \vspace{8cm}
  \caption{Plot showing the close correlation of $F_v(T_e,C)$ and
   $F_v(V_0,\phi)$ for the G-K dwarfs of Perrin \& Karoji (1987). The fitted
   straight line was obtained by an orthogonal linear regression, constrained
   to pass through the origin, $(0,0)$. The best-fit slope is consistent with
   unity and the correlation coefficient, $\rho = 0.95$, is highly
   significant.}
  \label{fig:fvcorrel}
\end{figure}
Following the argument in HOC we equate equation \ref{eq:logTeOHC} to equation
\ref{eq:FvOHC}, convert to reddened colour index and apparent magnitude and
make the angular diameter the subject of the equation thus:-
\begin{equation}
\label{eq:logphiOHC}
   \log{\phi} = 0.5054 - 0.082 \rmsub{E}{B - V} + 0.682 (B - V) - 0.2V
\end{equation}
where $\rmsub{E}{B - V}$ is the colour excess. Here we have assumed the
relation
between visual absorption ($A_v$) and colour excess to be given by
$A_v \simeq 3.0\rmsub{E}{B - V}$.

In the next section we apply equation \ref{eq:logphiOHC} to determine an
angular diameter estimate for each of our sampled stars, and then input these
estimates to our new method for measuring cluster distances.

\section{APPLICATION OF A NEW DISTANCE MEASURING TECHNIQUE TO REAL
STELLAR DATA}
\label{sec:applic}

Armed with our newly recalibrated Barnes-Evans relation, we are now in a
position to apply the distance method introduced in HOC to determine distances
to the Pleiades and $\alpha$ Persei clusters, from our cluster samples of
10 and 17 stars respectively.

\subsection{Angular diameters and projected cluster distances}
\label{sec:phidsini}

Using the recent photometric studies of the Pleiades by
\scite{stauffer87} and $\alpha$ Persei by \scite{stauffer85},
\scite{stauffer89} and \scite{prosser91} we obtained
accurate determinations of the colour index, $B - V$, mean apparent visual
magnitude, $\rmsub{V}{mean}$ and amplitude of photmetric modulation,
$\Delta V$, for
each of our sample stars. An average colour excess of 0.10 was adopted for
the $\alpha$ Persei region \cite{crawfordb74,prosser91}; for our Pleiades
sample we applied individual estimates of the colour excess to each star
\cite{stauffer87}. These data are summarised in table \ref{tab:logphidata}.

\begin{table*}
  \centering
  \caption{Photometric data and angular diameters, $\phi$,
($\times 10^{-5}$ arcsec) estimated from the recalibrated Barnes-Evans
relation for stars in the Pleiades and $\alpha$ Persei clusters.}
  \label{tab:logphidata}
  \input logphitab.tex
\end{table*}

Inserting these values into equation \ref{eq:logphiOHC} we obtained an
angular diameter
estimate, $\phi$, for each star: these estimates are also listed in table
\ref{tab:logphidata}. It is important to note that we do {\em not \/} insert
the mean apparent visual magnitude, $\rmsub{V}{mean}$, directly into equation
\ref{eq:logphiOHC}, but instead use a corrected value given by $\rmsub{V}{mean}
- \Delta V$. This is the apparent magnitude of the star at the point in its
rotational period when the fractional coverage of the visible hemisphere by
surface inhomogeneities is at a minimum: i.e. the point at which the
physical characteristics of the star (specifically its surface brightness and
effective temperature) most closely approximate those of an `immaculate
disk'. Although this correction is small, its
effect is systematic and thus a failure to account for it by using simply
$\rmsub{V}{mean}$ in equation \ref{eq:logphiOHC} would result in a small
positive bias in our cluster distance estimate.

We then combined these angular diameters with the measured rotational periods
and projected rotation velocities to derive a projected cluster distance
estimate, $D \sin i$, for each star, using equation \ref{eq:dsinival} above.
For our Pleiades sample the axial rotation periods were obtained from
\scite{lockwood84}, \scite{stauffer85}, \scite{leeuwen87},
\scite{magnitskii87}, \scite{stauffer88} and \scite{strassmeier88}. For
$\alpha$ Persei periods were obtained from \scite{stauffer85},
\scite{stauffer89}, \scite{prosser91}, \scite{prosser93}, \scite{odell93} and
\scite{odell93a}. Projected rotational velocities were obtained from
\scite{stauffer87} for the Pleiades stars, and from \scite{stauffer85},
\scite{stauffer89} and \scite{prosser91} for $\alpha$ Persei.
The results for both cluster samples are summarised in table
\ref{tab:dsinidata}, where the
samples have now been re-ordered by increasing $D \sin i$. Henceforth we
denote an {\em ordered \/} sample of $D \sin i$ values as $\{ D \sin i_{(k)}
\hspace{2mm} ; k = 1 , n \} $, where:-
\begin{equation}
\label{eq:dsiniord}
   D \sin i_{(1)} \leq D \sin i_{(2)} \leq \ldots \leq D \sin i_{(n)}
\end{equation}
and where evidently $n = 10$ and $n = 17$ for the Pleiades and $\alpha$ Persei
samples respectively.

\begin{table*}
  \centering
  \caption{Rotational data, angular diameters and projected cluster distance
           estimates, $D \sin i$, for stars in the
   Pleiades and $\alpha$ Persei clusters. (Periods denoted by an asterisk are
   preliminary results, to be confirmed in O'Dell et al. 1993).}
  \label{tab:dsinidata}
  \input dsinitab.tex
\end{table*}

\subsection{Modelling the distribution of $\alpha$}
\label{sec:adistrib}

Following the method described in HOC we must next model the distribution of
$D \sin i$
values {\em expected \/} in a cluster at a given true distance. In general our
model should incorporate: 1) the intrinsic joint distribution of rotation
velocity, rotation period, inclination and angular diameter; 2) the scatter
due to measurement error in the observed values of these variables; 3) the
observational selection effects to which the sample is subject. Recall from
HOC, however, that our model can be considerably simplified by introducing a
dimensionless variable, $\alpha$, defined by:-
\begin{equation}
\label{eq:alphadefn}
   \alpha \: \: =  \: \: \frac{\rmsub{(v \sin i)}{obs}}
                          {\rmsub{v}{true} \: z_{\phi}}
\end{equation}
Here $\rmsub{(v \sin i)}{obs} \,$ is the observed projected rotation velocity,
$\rmsub{v}{true} \,$ is the true equatorial rotation velocity and
$z_{\phi} \: = \: \frac{\phi_{obs}}{\phi_{true}}$ is a random variable defined
as the ratio of the `observed' angular diameter (i.e. the value inferred from
equation \ref{eq:logphiOHC}) to the true angular diameter of a given star.

We may write $D \sin i$ in terms of $\alpha$ as follows:-
\begin{equation}
\label{eq:dsinialpha}
   D \sin i  = \rmsub{D}{true} \: \alpha
\end{equation}
from which it is clear that we can obtain the $D \sin i$ distribution expected
for a cluster at any true distance simply by rescaling the distribution
of $\alpha$.

We derive the distribution of $\alpha$ via Monte Carlo simulations, as
illustrated in HOC. The Monte Carlo sampling was carried out as follows.

\begin{enumerate}
\item{$\rmsub{v}{true}$ was drawn from a uniform distribution over the range
0 to 240kms$^{-1}$. The upper limit of 240kms$^{-1}$ was adopted since it
corresponds to a rotation period of $\sim$ five hours for a star of one solar
radius. As can be seen from table (2), this is approximately equal to the
shortest measured rotation periods in both the Pleiades and $\alpha$ Persei
samples.}
\item{A true inclination, $i$, was assigned by sampling $\cos i$ from
the uniform distribution over $[ \, 0 \, , \, 1 \,] \,$ - i.e. assuming
that the rotation axis has no preferred direction in space
(\cite{Slettebak70}).}
\item{$\rmsub{(v \sin i)}{obs}$ was computed by multiplying
$\rmsub{(v \sin i)}{true}$ by a random measurement error drawn from a Gaussian
of unit mean and dispersion 0.1 - i.e. an observational scatter of $10 \%$.}
\item{A lower selection limit of $\rmsub{(v \sin i)}{obs} \geq 45$kms$^{-1}$
and $\rmsub{(v \sin i)}{obs} \geq 50$kms$^{-1}$ was then imposed for the
Pleiades and $\alpha$ Persei samples respectively. Any star failing to meet
this criterion was rejected from the sample and the Monte Carlo procedure
restarted. This selection function was primarily designed to select fast
rotators - for which rotation periods had already been, or could be, measured
photometrically within the available observing time \cite{odell93}. As we
discuss in more detail below, however, this lower limit also serves to exclude
stars of low inclination.}
\item{Finally, $z_{\phi}$ was drawn from a gaussian of unit mean and dispersion
0.1. Note that assigning a constant {\em percentage \/} error dispersion to the
observed angular diameter is consistent with a constant dispersion in the
relation for $\log \phi$ given by equation (7).}
\end{enumerate}

Figures \ref{fig:pdf_alpha_45} to \ref{fig:cdf_alpha_50} show probability
density (pdf) and cumulative distribution (cdf) curves for $\alpha$, modelled
for the Pleiades and $\alpha$ Persei samples, obtained by spline
fitting to histograms constructed from 50000 trials. Note that there is
essentially no difference between the shapes of these curves for the two
samples - indicating that the distribution of $\alpha$ is not very sensitive
to the adopted lower limit of $\rmsub{(v \sin i)}{obs}$.
\begin{figure}
  \centering
  \vspace{6cm}
  \caption{Spline fit to the probability density function of $\alpha$,
   modelled by 50000 Monte Carlo trials and assuming
   $\rmsub{(v \sin i)}{obs} \geq 45$kms$^{-1}$, corresponding to our sample
   of Pleiades stars.}
  \label{fig:pdf_alpha_45}
\end{figure}
\begin{figure}
  \centering
  \vspace{6cm}
  \caption{Spline fit to the cumulative distribution function of $\alpha$,
   modelled by 50000 Monte Carlo trials and assuming
   $\rmsub{(v \sin i)}{obs} \geq 45$kms$^{-1}$, corresponding to our sample
   of Pleiades stars.}
  \label{fig:cdf_alpha_45}
\end{figure}
\begin{figure}
  \centering
  \vspace{6cm}
  \caption{Spline fit to the probability density function of $\alpha$,
   modelled by 50000 Monte Carlo trials and assuming
   $\rmsub{(v \sin i)}{obs} \geq 50$kms$^{-1}$, corresponding to our sample
   of $\alpha$ Persei stars.}
  \label{fig:pdf_alpha_50}
\end{figure}
\begin{figure}
  \centering
  \vspace{6cm}
  \caption{Spline fit to the cumulative distribution function of $\alpha$,
   modelled by 50000 Monte Carlo trials and assuming
   $\rmsub{(v \sin i)}{obs} \geq 50$kms$^{-1}$, corresponding to our sample
   of $\alpha$ Persei stars.}
   \label{fig:cdf_alpha_50}
\end{figure}
As we noted above and discussed briefly in HOC, our lower selection limit
of $\rmsub{(v \sin i)}{obs}$ indirectly excludes stars of low inclination since
we are assuming a {\em maximum \/} rotation velocity -- in this case
$\rmsub{v}{true} = 240$kms$^{-1}$. Specifically, a limit of
$\rmsub{(v \sin i)}{obs} \geq 50$kms$^{-1}$ implies that $i \geq 12^{\circ} \,$
because a star of lower inclination,
even with the maximum rotation velocity of 240kms$^{-1}$, would have
$\rmsub{(v \sin i)}{obs} < 50$kms$^{-1}$ and thus would not be selected.
As we discuss in section \ref{sec:disc}, in reality the inclination selection
function may be somewhat more complex than this simple step function when
photometric selection effects are taken into account.
A more accurate model for the distribution of $\alpha$ may, therefore, demand
the inclusion of selection effects at higher inclinations. We will return to
this important issue in due course. For the moment, however, we will proceed
with the simple model developed thus far - i.e. where the inclination selection
is introduced solely as a result of our lower limit on $(v \sin i)_{obs} \,$.
We can regard the cluster distance estimated from this model as a useful
first approximation: indeed we will later use this approximation in
attempting to implement a more realistic inclination selection function.

\subsection{`Ordered' and Bayesian cluster distance estimators}
\label{sec:distests}

In HOC we introduced an `ordered' distance estimator, $\rmsub{\hat{D}}{ord}$,
\footnote{As in HOC, we adopt the usual statistical convention of denoting
an estimator of a parameter by a caret}
which is defined in terms of our ordered sample of $D \sin i$ estimates and
the pdf, $p_{(r)}$, of the order statistics of $\alpha$.
$\rmsub{\hat{D}}{ord}$ satisfies:-
\begin{equation}
\label{eq:dorddefn}
\frac{\partial \Lambda}{\partial D} | _{D = \rmsub{\hat{D}}{ord}}
\hspace{1mm} = \hspace{1mm} 0
\end{equation}
where $\Lambda$ is given by:-
\begin{equation}
\label{eq:Lambdadefn}
\Lambda_{r}(D) = \prod_{r=1}^{n} p_{r}(\alpha_{(r)} = \frac{D \sin i_{(r)}}{D}
)
\end{equation}
Thus $\rmsub{\hat{D}}{ord}$ is the distance which maximises the product over
the likelihoods of obtaining each $D \sin i_{(r)} \hspace{2mm} ; r = 1 , n$.
(See HOC for more details).

Expressing the pdfs of the order statistics of $\alpha$ in terms of the
appropriate modelled pdf and cdf, as given by figures \ref{fig:pdf_alpha_45}
to \ref{fig:cdf_alpha_50}, and substituting into equation
\ref{eq:Lambdadefn} the $D \sin i$ data for each cluster from table
\ref{tab:dsinidata}, we obtained ordered distance estimates of
$\rmsub{\hat{D}}{ord} = 134$pc for the Pleiades cluster and
$\rmsub{\hat{D}}{ord} = 195$pc for the $\alpha$ Persei cluster.

Error estimates were computed as described in HOC, by repeating our Monte
Carlo procedure on 2000 synthetic cluster samples for each real data set, at
an assumed distance of 134pc and 195pc respectively. The histograms obtained
are shown in figures \ref{fig:hist_plei} and \ref{fig:hist_aper}. We can see
from these figures that the distribution of $\rmsub{\hat{D}}{ord}$ shows no
significant departure from normality: thus we can adopt the dispersion as a
measure of the error on $\rmsub{\hat{D}}{ord}$ in the usual way. This gives
the following results (with $1 \, \sigma$ errors):-

\begin{description}
\item[Pleiades] : \hspace{4mm} 134 $\pm$ 10 pc
\item[$\alpha$ Persei] : \hspace{4mm} 195 $\pm$ 11 pc
\end{description}
\begin{figure}
  \centering
  \vspace{6cm}
  \caption{Histogram of ordered distance estimates, $\rmsub{\hat{D}}{ord}$,
   obtained from 2000 synthetic Pleiades cluster samples, at an assumed
distance
   of 134pc}
   \label{fig:hist_plei}
\end{figure}
\begin{figure}
  \centering
  \vspace{6cm}
  \caption{Histogram of ordered distance estimates, $\rmsub{\hat{D}}{ord}$,
   obtained from 2000 synthetic $\alpha$ Persei cluster samples, at an assumed
   distance of 195pc}
  \label{fig:hist_aper}
\end{figure}
We can also construct a Bayesian estimate of the cluster distance, following
the method described in HOC section (5). Figures \ref{fig:plei_post} and
\ref{fig:aper_post} show the posterior distribution for the true distance,
$\rmsub{D}{true}$, of the Pleiades and $\alpha$ Persei clusters - assuming
a uniform prior distribution. The results are clearly consistent with the
ordered distance estimates and their errors quoted above.
\begin{figure}
  \centering
  \vspace{6cm}
  \caption{Posterior distribution for the true cluster distance,
  $\rmsub{D}{true}$, derived by applying the bayesian method described in
  HOC to the Pleiades sample - assuming a uniform prior.}
  \label{fig:plei_post}
\end{figure}
\begin{figure}
  \centering
  \vspace{6cm}
  \caption{Posterior distribution for the true cluster distance,
  $\rmsub{D}{true}$, derived by applying the bayesian method described in
  HOC to the $\alpha$ Persei sample - assuming a uniform prior.}
  \label{fig:aper_post}
\end{figure}
\subsection{Improving the modelled distribution of $\alpha$}
\label{sec:abetter}

The distance estimate of 134pc for the Pleiades cluster is in excellent
agreement with recent results obtained from ZAMS fitting techniques (see
table \ref{tab:distdata} above). The value of 195pc obtained for $\alpha$
Persei, on the other hand, is considerably larger than existing ZAMS estimates.
In response to this discrepancy, in the next section we will apply the
recent semi-empirical main-sequence fitting technique of \scite{vandp89}
to the $\alpha$ Persei data in order to check the reliability of previous
ZAMS distance estimates.
Before we do so, however, we should first consider in more detail the main
source of systematic error in our cluster distance estimate for
$\alpha$ Persei: the limitations of our simple model
for the inclination selection function. We can demonstrate the limitations of
our model as follows. Suppose that the true distance of $\alpha$ Persei {\em
were \/} equal to 195pc. If we divide each $D \sin i_{(k)} \hspace{2mm} k =
1...n$, by 195pc we obtain a sample of the order statistics, $\alpha_{(k)}$,
of $\alpha$.
Using these order statistics we can construct a sample estimate of the cdf,
$\Phi (\alpha)$, of $\alpha$, given by:-
\begin{equation}
\label{eq:phi_alpha}
\Phi (\alpha) = \left\{ \begin{array}{ll}
               0  & \hspace{5mm}
               \mbox{} \alpha \leq \alpha_{(1)} \\
               \frac{i}{n}  & \hspace{5mm}
               \mbox{} \alpha_{(i)} < \alpha \leq \alpha_{(i+1)}
               \hspace{2mm} i = 1 \ldots n-1 \\
               1  & \hspace{5mm}
               \mbox{} \alpha > \alpha_{(n)} \\
                  \end{array}
                  \right.
\end{equation}
Hence $\Phi (\alpha)$ is just a step function incrementing  at the
values of the order statistics.

We can then compare $\Phi (\alpha)$ with the modelled cdf of $\alpha$ for the
$\alpha$ Persei sample, as shown in figure \ref{fig:cdf_alpha_50} above.
If our model for the inclination selection effects is a good one, then
our sample and model cdf should be in good agreement.

Figure \ref{fig:cdf_1_aper} shows this comparison for our $\alpha$ Persei
sample data, assuming a true cluster distance of 195pc. We can see that the
sample cdf deviates markedly from the model curve at both high and low values
of $\alpha$. This indicates that our simple model for the inclination selection
function is inadequate to describe accurately the inclination distribution
of the $\alpha$ Persei data, and this inadequacy undermines the cluster
distance estimate obtained with this model.

As a comparison figure \ref{fig:cdf_1_plei} shows the sample and model cdf
curves of $\alpha$ obtained from our Pleiades data, assuming a true cluster
distance of 134pc. We see that in this case considerably better agreement
between the curves is obtained over the full range of $\alpha$.
\begin{figure}
  \centering
  \vspace{6cm}
  \caption{Comparison between the sampled and modelled cdf of $\alpha$
   for the $\alpha$ Persei data, assuming a true cluster distance of 195pc}
  \label{fig:cdf_1_aper}
\end{figure}
\begin{figure}
  \centering
  \vspace{6cm}
  \caption{Comparison between the sampled and modelled cdf of $\alpha$
   for the Pleiades data, assuming a true cluster distance of 134pc}
  \label{fig:cdf_1_plei}
\end{figure}
These qualitative comparisons are borne out by more rigorous statistical
analysis using the Kolmogorov-Smirnov (KS) statistic, usually denoted
$\rmsub{D}{n}$. (c.f. \scite{KendallStuart63}). This statistic is defined as
the maximum absolute deviation between the sample and model cdf curves. We find
from figures \ref{fig:cdf_1_aper} and \ref{fig:cdf_1_plei} that
$\rmsub{D}{n} = 0.182$ and $\rmsub{D}{n} = 0.142$ for $\alpha$ Persei and
the Pleiades respectively: the larger value of $\rmsub{D}{n}$ for $\alpha$
Persei is more significant when one takes into account the larger sample size
for this cluster. The more relevant statistic in this case is the
{\em reduced \/} KS statistic, $\rmsub{Z}{n} = n^{\frac{1}{2}} \rmsub{D}{n}$.
We find that $\rmsub{Z}{n} = 0.752$ for $\alpha$ Persei and $\rmsub{Z}{n} =
0.448$ for the Pleiades.

It seems clear, therefore, that in order to improve our cluster distance
estimate for $\alpha$ Persei we must introduce a more realistic model for
the sample inclination selection. The precise form of this selection function
at higher inclinations will, in general, depend closely upon the distribution
and composition of surface features on the sampled stars. In a fully rigorous
treatment one could begin by modelling this surface distribution and deduce
from it the probability, as a function of inclination, of detecting rotational
modulation of a given amplitude. Clearly in such an approach the inclination
selection function would be more complex in form than the simple step function
assumed above. We will consider an analysis of this kind in our next paper.
For the present, however, we can improve considerably upon our model for the
inclination selection while still assuming it to take the form of a step
function simply by increasing the lower inclination limit, $\rmsub{i}{lim}$.
Moreover, rather than adopting some {\em ad hoc \/} value for $\rmsub{i}{lim}$,
we can motivate our choice using the data samples themselves. We proceed as
follows.

\begin{enumerate}
\item{Assuming a range of different values for $\rmsub{i}{lim}$ we derive
the distribution of $\alpha$ for each $\rmsub{i}{lim}$ following the same
Monte Carlo procedure as before - save only an additional step to verify
that in each Monte Carlo trial $\rmsub{i}{true} \geq \rmsub{i}{lim}$.}
\item{For each $\alpha = \alpha(\rmsub{i}{lim})$ distribution we compute an
ordered cluster distance estimate from the real cluster data, applying
equations \ref{eq:dorddefn} and \ref{eq:Lambdadefn} as before.}
\item{Assuming that the true cluster distance is equal to the ordered distance
estimate we next compute a sample cdf for $\alpha$ and compare it with the
modelled distribution for each inclination limit, calculating the KS statistic
for the two curves in each case.}
\item{Finally, we adopt as our `best-fit' inclination limit the value of
$\rmsub{i}{lim}$ which minimises the KS statistic for $\alpha$, and we adopt
as our cluster distance estimate the value of $\rmsub{\hat{D}}{ord}$ which we
obtain when we assume this inclination limit.}
\end{enumerate}

We applied the above procedure to both our Pleiades and $\alpha$ Persei
samples, incrementing $\rmsub{i}{lim}$ at $1^{\circ}$ intervals between
$0^{\circ}$ and $80^{\circ}$. For the $\alpha$ Persei data we obtained a
minimum KS statistic of $\rmsub{D}{n} = 0.126$ when $\rmsub{i}{lim} =
41^{\circ}$: a reduction of more than 25\% compared with $\rmsub{D}{n}$ for
figure \ref{fig:cdf_1_aper}. This new inclination limit yielded a distance
estimate of $\rmsub{\hat{D}}{ord} = 187$pc. Figure \ref{fig:cdf_2_aper}
compares the sample and model cdf curves of $\alpha$, assuming a true distance
of 187pc and $\rmsub{i}{lim} = 41^{\circ}$, and clearly indicates that the
data provide a much better fit to this improved model for the inclination
selection.
\begin{figure}
  \centering
  \vspace{6cm}
  \caption{Comparison between the sampled and modelled cdf of $\alpha$
   for the $\alpha$ Persei data, assuming a true cluster distance of 187pc,
   corresponding to the `best-fit' lower inclination limit of $\rmsub{i}{L}
   = 41^{\circ}$.}
  \label{fig:cdf_2_aper}
\end{figure}
For the Pleiades data we obtained a minimum KS value of $\rmsub{D}{n} = 0.120$
for $\rmsub{i}{lim} = 23^{\circ}$. This limit gave a cluster distance
estimate of $\rmsub{\hat{D}}{ord} = 132$pc. Figure \ref{fig:cdf_2_plei}
compares the sample and model cdf curves for the Pleiades data assuming this
revised distance and inclination limit. It is not surprising that we found
little improvement in the value of $\rmsub{D}{n}$ for the Pleiades sample since
the data already gave a good fit to our previous simple model.
Consequently the improved distance estimate for the Pleiades is virtually
identical to our first approximation.
\begin{figure}
  \centering
  \vspace{6cm}
  \caption{Comparison between the sampled and modelled cdf of $\alpha$
   for the Pleiades data, assuming a true cluster distance of 132pc,
   corresponding to the `best-fit' lower inclination limit of $\rmsub{i}{L}
   = 23^{\circ}$.}
  \label{fig:cdf_2_plei}
\end{figure}
We next recomputed error estimates for each cluster distance by repeating
our Monte Carlo procedure on 2000 synthetic cluster samples for each real
data set - now assuming the appropriate revised true distance and inclination
limit. No change was found in the dispersion of the histograms which we
obtained. Thus we adopt as the distance (modulus) estimate to each cluster
using our new method the following values (with 1 $\sigma$ errors):-

\begin{description}
\item[Pleiades] : \hspace{4mm} 132 pc $\pm$ 10 \hspace{6mm}
(5.60 mag. $\pm$ 0.16)
\item[$\alpha$ Persei] : \hspace{4mm} 187 pc $\pm$ 11 \hspace{6mm}
(6.35 mag. $\pm$ 0.13)
\end{description}

Finally we constructed a Bayesian estimate of each cluster distance, now
assuming the appropriate lower inclination limit. Figures
\ref{fig:post_2_plei} and \ref{fig:post_2_aper} show the posterior distribution
for the true distance,
$\rmsub{D}{true}$, of the Pleiades and $\alpha$ Persei clusters respectively
- assuming a uniform prior distribution. Again the results are consistent
with our revised ordered distance estimates.
\begin{figure}
  \centering
  \vspace{6cm}
  \caption{Posterior distribution for the true cluster distance,
   $\rmsub{D}{true}$, assuming a lower inclination limit of
   $\rmsub{i}{lim} = 23^{\circ}$, derived from the Pleiades sample.}
  \label{fig:post_2_plei}
\end{figure}
\begin{figure}
  \centering
  \vspace{6cm}
  \caption{Posterior distribution for the true cluster distance,
   $\rmsub{D}{true}$, assuming a lower inclination limit of
   $\rmsub{i}{lim} = 41^{\circ}$, derived from the $\alpha$ Persei sample.}
   \label{fig:post_2_aper}
\end{figure}
\section{APPLICATION OF A ZAMS FITTING TECHNIQUE TO ALPHA PERSEI}
\label{sec:zams}

In the light of the results of the previous section (notably the
distance to $\alpha$ Persei) we considered it
pertinent to compare our estimates with those of another technique,
one which relies on a completely different set of model assumptions.

Reviewing the current technique of distance measurement for nearby star
clusters we concluded that the ZAMS fitting method of \scite{vandp89},
(hereafter VP) was one of the the most accurate alternatives available.

The precursor of this work was the paper by \scite{vandb84} who
modelled theoretical zero-age main-sequences based on predetermined
helium and metal abundances ($Y$ and $Z$ respectively) and mixing length
ratios ($\alpha$). Transposing these ZAMS to the observational
($M_v$, $B-V$) plane allowed them to make a direct comparison
with the observations of the nearby Pleiades and Praesepe star clusters.
An accurate fit of the theoretical ZAMS to the observational data
placed constraints on $Y$, $Z$ and $\alpha$ enabling a `preferred' ZAMS
locus to be found. This was then applied to the Hyades C-M diagram to
obtain a distance modulus for the cluster.
Their result highlighted the problem that the lack of a well-defined zero
point for the colour scale can lead to considerable uncertainty
in the derived distance -- specifically because the method requires an
accurate $B-V$ colour for the Sun (as in fact is the case with all
ZAMS fitting techniques).

The work of VP refined and expanded this approach by reducing
the uncertainty in the zero point calibration and creating a
semi-empirical main-sequence fitting technique which we now briefly
review before applying it to our cluster data.

In the development of their method
they employed the Hyades cluster as a distance calibrator and the
Pleiades as an indicator of the morphological profile of a ZAMS
with solar-type abundances.
Firstly they circumvented the initial requirement of an accurate
knowledge of the solar $B-V$ by instead
normalising their ZAMS models in the ($M_v$, $T_{eff}$)
plane, using the accurate effective temperatures of a total 19 Hyades
dwarfs derived by \scite{duncanj83} and \scite{cayrelc85}.
VP then adopted for the Hyades the ZAMS with predetermined parameters
$Y$ = 0.27 (i.e $Y = Y_{\odot}$, assuming a near solar helium
content) and metallicity, $Z$ = 0.0240 (corresponding to [Fe/H] = 0.15).
They next matched this theoretical ZAMS to the Hyades observations plotted
on the ($M_v$, $T_{eff}$) plane, assuming the cluster distance.
It is important to note that
VP fitted the ZAMS to a selected population of the cluster, namely those
members that lie close to their respective main-sequence `ridge' lines
(these being assumed central cluster members).
Having obtained a good fit in the ($M_v$, $T_{eff}$) plane they showed
that it is possible to obtain consistent results in the ($M_v$, $B-V$)
plane for the same distance providing a shift of 0.03 mag. is applied
redward to the colour axis. This plot implies the Sun's colour to be
$(B-V)_{\odot}$ = 0.63 - 0.64, which is consistent with other estimates.

Having normalised the colours and luminosities to the theoretical ZAMS
using the Hyades, VP then considered changes in the helium and metal
abundances and their effect when removed from the solar values.
In the present context we need not consider a change in the helium
abundance from the solar value, $Y$ = 0.27, since the age of
$\alpha$ Persei is comparable with that of the Pleiades.
With regard to the metallicity however we considered it important
to apply the correction given by VP, using accurate
values of [Fe/H] otained from a study of the
chemical composition of open clusters by \scite{boesgaardf90}. VP
established the metallicity dependence of their ZAMS models using the
population II subdwarf, Groombridge 1830, through which they calibrated
the low metallicity end of the $M_v$/[Fe/H] relation. They
discovered that the best fit to this relation was described by the
following parabolic curve:
\begin{equation}
\label{eq:[Fe/H]}
  {\delta}{M_v}(Fe/H) = -[Fe/H](1.444 + 0.362[Fe/H]
\end{equation}
This relation gives the resultant shift in magnitude at a fixed
$B-V$ between the solar abundance ZAMS and one of a different
metallicity.

Having established how $M_v$ will vary due to changes in composition,
VP considered how evolutionary changes shift the position of a star in
relation to the ZAMS.
They succeeded in modelling this shift by fitting a polynomial to the
morphological profile of the Pleiades ZAMS on the C-M diagram.
Adjusting the constant term of this equation so as to normalise to the
Sun they obtained the following expression:
\begin{eqnarray}
\label{eq:B-V}
  M_v(B-V) & = & 2.836 - 6.796(B-V) + 31.77(B-V)^2 - \nonumber \\
           &   & 31.6(B-V)^3 + 10.57(B-V)^4
\end{eqnarray}
This quartic is valid over the colour range $0.2{\leq}B-V{\leq}1.0$ and
determines the location of the ZAMS for stars with solar abundances,
i.e $Y$ = 0.27 and [Fe/H] = 0.

Assuming solar composition for the Pleiades and $\alpha$ Persei we need
now only to introduce equation \ref{eq:[Fe/H]} to adjust for their specific
metallicities. To this end VP therefore defined the following equation
(we have excluded the $Y$ term which for our clusters is zero):
\begin{equation}
\label{eq:V}
  V = {M_v}(B-V) + {\delta}{M_v}(Fe/H) + (m-M)_v
\end{equation}
It is now a simple matter of subtracting the metallicity-corrected
magnitude on the ZAMS from that of the star
to obtain the distance modulus $(m - M)_v$.

The above technique has been applied by VP to a number of open
clusters. For the Pleiades they obtained a distance modulus,
$(m - M)_v$ = 5.60 mag. assuming zero reddening
($E_{B-V} = 0.0$) and [Fe/H] = 0.05, thus confirming the estimate
in the earlier paper by \scite{vandb84},
(see table \ref{tab:distdata}.)
This result is in excellent agreement with the distance estimated using
our new method in section \ref{sec:applic} above.

We now apply the ZAMS fitting technique of VP to derive a distance
modulus for $\alpha$ Persei.

The photometry and reddening data for all the known cluster members
were extracted from the work of \scite{mitchell60}, \scite{crawfordb74},
\scite{stauffer85,stauffer89} and \scite{prosser91}.
\scite{crawfordb74} gave individual reddening corrections for a large
number of the brighter members of the cluster in the Str{\"{o}}mgren
photometric system. These values lie within the range,
$E_{b-y} = 0.03 - 0.16$ mag. Following the standard reddening law
of \scite{divan54} and \scite{whitford58}, (see also \scite{stromgren63}),
we converted to the Johnson system using:
\begin{equation}
\label{eq:Eb-y}
  E_{b-y} = 0.70E_{B-V}
\end{equation}
Hence the colour excess for the $\alpha$ Persei cluster lies within the
range, $E_{B-V} = 0.04 - 0.23$ mag. with an average value of
$E_{B-V} = 0.10$ mag. which is confirmed by \scite{prosser91}.
The members listed by \scite{crawfordb74} were corrected
individually whilst the remainder were assigned the average reddening
of the cluster.
In addition we applied a visual absorption correction $(A_v)$
of 0.30 mag. derived
by assuming a value of 3.0 for the ratio of total to selective absorption
in the $B-V$ system, as determined by \scite{hiltner56} for the
Perseus region.

In addition to the reddening and absorption corrections we applied
a metallicity correction using equation \ref{eq:[Fe/H]} with
$[Fe/H] = 0.05$. This value is an approximation of that
given in \scite{boesgaardf90}.

Finally, after constructing the C-M diagram of the cluster where -
plotting unreddened apparent magnitude against unreddened colour - we
fitted the quartic given by equation \ref{eq:B-V} to the
supposed ZAMS of the cluster population.
The fit was carried out by shifting the quartic vertically through
the C-M diagram until it reached the `ridge line', or supposed ZAMS of
the cluster. At each vertical position the distance to the nearest
star was summed over all points along the quartic, and the `ridge line'
was identified at the vertical position which minimised this sum -
thus locating the highest density region of the C-M diagram. This
`best-fit' vertical displacement defined the cluster distance modulus,
following equation \ref{eq:V}.

Our best fit correlated well with an earlier fit by
\scite{crawfordb74} using the well known ZAMS calibration of
\scite{blaauw63}.
Both fits are shown in figure \ref{fig:CM_diag} where a distance
$(m - M)_v = 6.25$ mag. is recovered.
\begin{figure*}
  \centering
  \vbox to100mm{}
  \caption{Colour-magnitude diagram showing a main-sequence fit to the
   $\alpha$ Persei cluster usinq the technique of Vandenberg and
   Poll(1989) for $m - M_v$ = 6.25 mag. The solid curve is a quartic valid
   only for $0.2{\leq}B-V{\leq}1.0$ where $Y$ = 0.27 and [Fe/H] = 0.05.
   The diagram has been corrected for reddening.}
  \label{fig:CM_diag}
\end{figure*}
In the next section we discuss the
implications of this result which, combined with our estimate in
section \ref{sec:applic}, supports the possibility that $\alpha$ Persei
is more distant than recently supposed.

\section{DISCUSSION}
\label{sec:disc}

For comparison the results of section \ref{sec:applic} and \ref{sec:zams}
are summarised in Table \ref{tab:distsumm}.

\begin{table*}
  \centering
  \caption{ Distances for the Pleiades and $\alpha$
   Persei open clusters determined by the main-sequence fitting method
   of Vandenberg and Poll(1989), (VP), and the statistical method of
   Hendry, O'Dell and Cameron(1993), (HOC).}
  \label{tab:distsumm}
  \input dist2-tab.tex
\end{table*}

Both methods give distance estimates which are in good agreement, the ZAMS
distance to each cluster lying within the 1 $\sigma$ error bar of the
distance estimate computed using our new method, as quoted in section
\ref{sec:applic}.

The excellent agreement between the Pleiades distance determinations lends
credibility to the basic principle of our new technique. The small
difference in the $\alpha$ Persei distances, on the other hand -- while
not statistically significant -- demonstrates that there is some scope
for improving our model for the observational selection effects to which the
sample is subject.

Of course the incompleteness of our model was already apparent in the results
of section \ref{sec:applic}. As we remarked in that section, the
sample cumulative distribution
of $D \sin i$ values for the Pleiades stars matches the modelled cdf curve
very well - irrespective of whether we impose a sharp inclination selection
limit independently of the limit on $\rmsub{(v \sin i)}{obs}$
(c.f. figures \ref{fig:cdf_1_plei} and \ref{fig:cdf_2_plei}).
In contrast, we can see from figures \ref{fig:cdf_1_aper} and
\ref{fig:cdf_2_aper} that the $\alpha$ Persei sample cumulative distribution
of $D \sin i$ values displays a small systematic deviation from the modelled
curve in both cases. Imposing a higher inclination limit clearly improves the
overall fit to the sample cdf -- as is indicated by the reduction in the value
of the KS statistic for figure \ref{fig:cdf_2_aper} -- but the match is
evidently not perfect, particularly in the tails of the distribution
($\alpha \lsim 0.7$, $\alpha \gsim 1.1$). The main reason for this discrepancy
is clear from table \ref{tab:dsinidata}, in which we see that the sampled
$D \sin i$ values are not uniformly spread, but rather cluster sharply
between 180 and 190pc. This `clumping' of the
$\alpha$ Persei data causes the sample cdf to rise sharply over the
corresponding narrow range of $\alpha$, as can be seen in both figures
\ref{fig:cdf_1_aper} and \ref{fig:cdf_2_aper}. It is possible to reproduce this
sharp rise in the cdf by increasing the inclination selection limit still
further -- to, say, $\rmsub{i}{lim} \simeq 60^{\circ}$ -- but
only at the cost of severely underestimating the low $\alpha$ tail of the
sample distribution. In other words one cannot, with our simple model for the
inclination selection, entirely explain both the clumping of $D \sin i$ values
close to 190pc {\em and \/} the considerably smaller values obtained for
HE699 and AP193.

Now, of course, it is possible that we {\em have} accurately modelled the
underlying inclination selection function for all stars in the $\alpha$
Persei cluster, and that our imperfect fit in figures \ref{fig:cdf_2_aper} is
due simply to the relatively small size of our sample.
One might therefore expect to obtain a more precise fit -- and obviously a
better distance estimate -- from a larger sample of stars. This however is
dependent on the acquisition of more axial rotation periods: a lengthy and
time consuming process!

A far more likely explanation for the discrepancy in figure
\ref{fig:cdf_2_aper} is that our model for the underlying inclination selection
function requires further improvement -- in particular the assumption of a
functional form which is more realistic than a simple step function.
Nevertheless, it seems clear from the distribution of $D \sin i$
values for $\alpha$ Persei, that a more realistic model would still
preferentially select stars of high inclination.

This behaviour provides some insight into the surface distribution of spots
on our $\alpha$ Persei stars. We found the best fit lower inclination limit
for this sample to be $\rmsub{i}{lim} = 41^{\circ}$.
If significant features were present only at equatorial latitudes then one
might expect that limb darkening would
prevent their detection in stars of very low inclination (i.e. $i \leq
10^{\circ}$). However, this argument would not be valid for
stars with inclinations in the range $\sim 10^{\circ} < i < 41^{\circ}$.
Consequently, the most likely physical explanation for the best fit lower
inclination limit -- and in particular the absence of stars with inclinations
in the above range -- is that spot activity is concentrated at high latitudes
in the rapid rotators, so that for $i \leq 40^{\circ}$ the change in the cosine
of the foreshortening angle reduces the amplitude of the spot modulation to
below our detection limit. This is consistent with the results of recent
Doppler
imaging studies (\scite{Strassmeier93}, \scite{Cameron93} and
\scite{hatzes92eieri}).

It is interesting to note that the considerably lower best fit inclination
limit for the Pleiades sample provides some evidence that the spot
distribution extends to lower latitudes in these stars.
The Pleiades is somewhat older than $\alpha$ Persei and its fast rotator
population has a substantially lower mean rotation velocity. It is therefore
conceivable that high latitude surface features are correlated with rapid
rotation rate, and hence with the age of the cluster.

The above arguments demonstrate clearly how the inclination selection
function depends strongly upon the distribution of surface features
on our sampled stars. Improving this aspect of our model
would not only increase the accuracy of our new distance
indicator, but would also improve our understanding of stellar magnetic
fields and surface activity in newly evolved late-type stars.

In conclusion we are satisfied that the Pleiades result validates our
method as a viable distance indicator.
Moreover, we believe that the distance of $\alpha$ Persei obtained with our
new indicator is supported by the results of our ZAMS fitting technique, and
that this distance is indeed further than other recent distance estimates to
the cluster.

\section{SUMMARY}
\label{sec:summ}

In this paper we have determined the distance to the Pleiades
and $\alpha$ Persei open
clusters using a new statistical distance indicator.
We find the distance modulus of the Pleiades to be
$(m - M)_v$ = 5.60 mag. $\pm$ 0.16, which corresponds extremely well
with the estimate obtained by a relatively recent ZAMS fitting
technique.
We find the distance modulus of $\alpha$ Persei
to be considerably larger than other recent estimates.
Our new distance indicator yields a distance modulus of
$(m - M)_v$ = 6.35 mag. $\pm$ 0.13. This estimate is
corroborated by ZAMS fitting, which indicates a distance modulus
of $(m - M)_v$ = 6.25 mag.

These results indicate that our new distance method, although limited in
application to young open clusters containing fast-rotating stars,
represents an important and useful addition to the range of techniques
for calibrating the local distance scale.

\vspace{5mm}
{\sc Acknowledgements} \\
 \\
The authors would like to thank Dave Clarke and Peter Panagi for their
helpful discussion.
We also acknowledge the use of the {\em Starlink \/} Microvax 3400
computer funded by a SERC grant to the Astronomy Centre at the
University of Sussex.
During the course of this work MAO was funded by a SERC research
studentship, MAH was supported by a SERC research fellowship and ACC
was supported by a SERC advanced fellowship at the University of Sussex.


\begin{thebibliography}{{Cayrel, Cayrel~de Strobel \& Campbell}{1985}}

\bibitem[\protect\citefmt{Barnes \& Evans}{1976}]{barevans76}
Barnes~T.~G., Evans~D.~S., 1976, MNRAS, 174, 489

\bibitem[\protect\citefmt{Barnes, Evans \& Moffett}{1978}]{barnes78}
Barnes~T.~G., Evans~D.~S., Moffett~T.~J., 1978, MNRAS, 183, 298

\bibitem[\protect\citefmt{Barnes, Evans \& Parsons}{1976}]{barnespar76}
Barnes~T.~G., Evans~D.~S., Parsons~S.~B., 1976, MNRAS, 174, 503

\bibitem[\protect\citefmt{Bernacca}{1970}]{Slettebak70}
Bernacca~P.~L., 1970, in Slettebak~A., ed, Stellar Rotation - Proceedings of
  the IAU colloquium held at the Ohio State University, U.S.A, 1969.
\newblock D. Reidel, Dordrecht-Holland, p.~227

\bibitem[\protect\citefmt{Blaauw}{1963}]{blaauw63}
Blaauw~A., 1963, in Strand~K.~A., ed, Basic Astronomical Data.
\newblock U, Chicago Press, Chicago, p.~383

\bibitem[\protect\citefmt{Boesgaard \& Friel}{1990}]{boesgaardf90}
Boesgaard~A.~M., Friel~E.~D., 1990, ApJ, 351, 467

\bibitem[\protect\citefmt{Cayrel, Cayrel~de Strobel \&
  Campbell}{1985}]{cayrelc85}
Cayrel~R., Cayrel~de Strobel~G., Campbell~B., 1985, A\&A, 146, 249

\bibitem[\protect\citefmt{Collier~Cameron \& Unruh}{1993}]{Cameron93}
Collier~Cameron~A.~C., Unruh~Y.~C., 1993, MNRAS

\bibitem[\protect\citefmt{Crawford \& Barnes}{1974}]{crawfordb74}
Crawford~D.~L., Barnes~J.~V., 1974, AJ, 79, 687

\bibitem[\protect\citefmt{Divan}{1954}]{divan54}
Divan~L., 1954, Ann. d'ap., 17, 456

\bibitem[\protect\citefmt{Duncan \& Jones}{1983}]{duncanj83}
Duncan~D.~K., Jones~B.~F., 1983, ApJ, 271, 663

\bibitem[\protect\citefmt{Gray}{1967}]{gray67}
Gray~D.~F., 1967, ApJ, 149, 317

\bibitem[\protect\citefmt{Gray}{1992}]{gray92}
Gray~D.~F., 1992, The observation and analysis of stellar photospheres, 2ed.
\newblock Cambridge University Press

\bibitem[\protect\citefmt{Hatzes \& Vogt}{1992}]{hatzes92eieri}
Hatzes~A.~P., Vogt~S.~S., 1992, MNRAS, 258, 387

\bibitem[\protect\citefmt{Hendry, O'Dell \& Collier~Cameron}{1993}]{hendry93}
Hendry~M.~A., O'Dell~M.~A., Collier~Cameron~A.~C., 1993, MNRAS, In press

\bibitem[\protect\citefmt{Hiltner \& Johnson}{1956}]{hiltner56}
Hiltner~W.~A., Johnson~H.~L., 1956, ApJ, 124, 367

\bibitem[\protect\citefmt{Johnson \& Hiltner}{1956}]{johnsonh56}
Johnson~H.~L., Hiltner~W.~A., 1956, ApJ, 123, 267

\bibitem[\protect\citefmt{Johnson \& Iriarte}{1958}]{johnsoni58}
Johnson~H.~L., Iriarte~B., 1958, Lowell Obs. Bull., 4, 47

\bibitem[\protect\citefmt{Johnson}{1957}]{johnson57}
Johnson~H.~L., 1957, ApJ, 126, 121

\bibitem[\protect\citefmt{Kendall \& Stuart}{1963}]{KendallStuart63}
Kendall~M., Stuart~A., 1963, The Advanced Theory of Statistics, vols I and II.
\newblock Haffner Publ. Co., New York

\bibitem[\protect\citefmt{Lockwood {\rm et~al.}}{1984}]{lockwood84}
Lockwood~G.~W., Thomson~D., Radick~R., Osborn~W., Baggett~W., Duncan~D.,
  Hartmann~L., 1984, PASP, 96, 714

\bibitem[\protect\citefmt{Magnitskii}{1987}]{magnitskii87}
Magnitskii~A.~K., 1987, SvA, 31, 696

\bibitem[\protect\citefmt{Mitchell}{1960}]{mitchell60}
Mitchell~R.~I., 1960, ApJ, 132, 68

\bibitem[\protect\citefmt{O'Dell \& Collier~Cameron}{1993}]{odell93}
O'Dell~M.~A., Collier~Cameron~A.~C., 1993, MNRAS, 262, 521

\bibitem[\protect\citefmt{O'Dell {\rm et~al.}}{1993}]{odell93a}
O'Dell~M.~A., Collier~Cameron~A.~C., Hilditch~R.~W., Bell~S.~A., 1993, in
  preparation

\bibitem[\protect\citefmt{Perrin \& Karoji}{1987}]{perrink87}
Perrin~M.-N., Karoji~H., 1987, A\&A, 172, 235

\bibitem[\protect\citefmt{Popper}{1980}]{popper80}
Popper~D.~M., 1980, ARA\&A, 18, 115

\bibitem[\protect\citefmt{Prosser {\rm et~al.}}{1993}]{prosser93}
Prosser~C.~F., Schild~R.~E., Stauffer~J.~R., Jones~B.~F., 1993, PASP, 105, 269

\bibitem[\protect\citefmt{Prosser}{1991}]{prosser91}
Prosser~C., 1991, {\rm PhD thesis}, University of California, Santa Cruz

\bibitem[\protect\citefmt{Sandage}{1957}]{sandage57}
Sandage~A., 1957, ApJ, 125, 435

\bibitem[\protect\citefmt{Stauffer \& Hartmann}{1987}]{stauffer87}
Stauffer~J.~R., Hartmann~L.~W., 1987, ApJ, 318, 337

\bibitem[\protect\citefmt{Stauffer, Hartmann \& Jones}{1989}]{stauffer89}
Stauffer~J.~R., Hartmann~L.~W., Jones~B.~F., 1989, ApJ, 346, 160

\bibitem[\protect\citefmt{Stauffer {\rm et~al.}}{1985}]{stauffer85}
Stauffer~J.~R., Hartmann~L.~W., Burnham~N., Jones~B.~F., 1985, ApJ, 289, 247

\bibitem[\protect\citefmt{Stauffer {\rm et~al.}}{1988}]{stauffer88}
Stauffer~J.~R., Schild~R., Baliunas~S.~L., Africano~J.~L., 1988, PASP, 99, 471

\bibitem[\protect\citefmt{Strassmeier {\rm et~al.}}{1988}]{strassmeier88}
Strassmeier~K.~G., Hall~D.~S., Zeilik~M., Nelson~E., Eker~Z., Fekel~F.~C.,
  1988, A\&AS, 72, 291

\bibitem[\protect\citefmt{Strassmeier {\rm et~al.}}{1993}]{Strassmeier93}
Strassmeier~K.~G., Rice~J.~B., Wehlau~W.~H., Hill~G.~M., Matthews~J.~M., 1993,
  A\&A, 268, 671

\bibitem[\protect\citefmt{Str{\"{o}}mgren}{1963}]{stromgren63}
Str{\"{o}}mgren~B., 1963, in Strand~K.~A., ed, Basic Astronomical Data.
\newblock U, Chicago Press, Chicago, p.~123

\bibitem[\protect\citefmt{van Leeuwen, Alphenaar \& Meys}{1987}]{leeuwen87}
van Leeuwen~F., Alphenaar~P., Meys~J. M.~M., 1987, A\&AS, 67, 483

\bibitem[\protect\citefmt{Vandenberg \& Bridges}{1984}]{vandb84}
Vandenberg~D.~A., Bridges~T.~J., 1984, ApJ, 278, 679

\bibitem[\protect\citefmt{Vandenberg \& Poll}{1989}]{vandp89}
Vandenberg~D.~A., Poll~H.~E., 1989, AJ, 98, 1451

\bibitem[\protect\citefmt{Whitford}{1958}]{whitford58}
Whitford~A.~E., 1958, AJ, 63, 201
\end{thebibliography}
\end{document}